\documentclass[a4paper,11pt]{article}
\pdfoutput=1 
\usepackage{jheppub} 
\usepackage[T1]{fontenc} 
\usepackage{bm,amssymb,slashed,graphicx,multirow,soul,mathtools,xspace,array,physics,xspace}
\usepackage[capitalise]{cleveref}  
\usepackage{float}   
\allowdisplaybreaks
\usepackage{ bbold }  
\usepackage{ifthen}
\usepackage{subcaption}
\usepackage{booktabs}
\usepackage[normalem]{ulem}
\usepackage[labelfont=bf]{caption}

\usepackage{tikz}
\usetikzlibrary{arrows,shapes}
\usetikzlibrary{trees}
\usetikzlibrary{matrix,arrows} 

\graphicspath{{./figs/}}  
\usetikzlibrary{patterns}

\newcommand{\HNLs}{RHNs\xspace}
\newcommand{\HNL}{RHN\xspace}
\newcommand{\vEW}{{\rm v}}
\newcommand{\mN}{{M_N}}
\newcommand {\e}{\mathrm{e}}
\newcommand {\iu}{\mathrm{i}} 

\title{\boldmath Coleman-Weinberg dynamics of ultralight scalar  \\ dark matter and GeV-scale right-handed neutrinos 
}

\author[a]{Clara Murgui,}
\author[a]{Ryan Plestid}

\affiliation[a]{Walter Burke Institute for Theoretical Physics, California Institute of Technology, Pasadena, CA 91125, USA}
\emailAdd{cmurgui@caltech.edu}
\emailAdd{rplestid@caltech.edu}

\abstract{We consider an extension of the Standard Model  by three singlet fermions and one singlet real scalar field. The scalar is an ultralight dark matter candidate whose abundance is set by dynamically induced misalignment from the Higgs portal. We focus on parameter space where the Coleman-Weinberg potential both fixes the dark matter relic abundance, and predicts the mass scale of right-handed neutrinos. The model prefers scalar masses in the range of  $10~{\mu\rm eV}\lesssim m_\phi\lesssim 10~{\rm meV}$, and can be tested via direct searches for a light scalar (e.g.\ fifth force tests), or by searching for right-handed neutrinos in laboratory experiments. }
\begin{document}

\maketitle

\section{Introduction \label{Intro}}

The origins of dark matter, neutrino masses, and the observed baryon asymmetry are three of the most important unanswered questions in fundamental physics. It is therefore interesting to consider extensions of the Standard Model (SM) that are capable of successfully addressing all three of these mysteries simultaneously. In this work we consider a (nearly) minimal extension of the SM by four singlet fields: one real scalar, and three right-handed neutrinos (\HNLs). The real scalar plays a two-fold role: it serves as a dark matter candidate and generates Majorana masses for the neutrinos.

In the absence of \HNLs our model reduces to the model suggested in \cite{Piazza:2010ye} in which a light scalar field couples dominantly via the super-renormalizable portal to the Higgs field $\mathcal{L}\supset -A \phi |H|^2$. One advantage of this model is that the scalar mass is protected from large radiative corrections proportional to the Higgs mass.\!\footnote{We will think of our model as an effective theory, valid below the electroweak scale, and focus only on radiative corrections induced by explicit degrees of freedom below the electroweak scale. We do not consider the coupling of $\phi$ to hypothetical heavy states above the electroweak scale. \label{footnote1}} This remains true provided that the renormalizable (i.e.\ quartic) Higgs-portal coupling takes on small values, not larger than the squared ratio of the light scalar mass to the Higgs mass. Nevertheless, even with a very soft coupling, i.e.\ $A \lesssim 1~{\rm \mu eV}$, the dark matter is not entirely secluded and interesting phenomenology still occurs, such that the model is testable and falsifiable. More recently, it has been noted that if the initial field misalignment after inflation is sufficiently small, then thermal misalignment \cite{Buchmuller:2004xr,Batell:2021ofv}  dominates such that the dark matter dynamics are insensitive to the initial conditions \cite{Batell:2021ofv,Batell:2022qvr}. This supplies a relic density target in close analogy with models of freeze-out dark matter, albeit with a very different microphysical origin.

The addition of \HNLs, or other physics capable of reproducing the observed neutrino textures \cite{deGouvea:2016qpx}, can qualitative change the scenario sketched above. Fermions coupled to scalars can have non-trivial thermal dynamics, see e.g.\  \cite{Lillard:2018zts}. Perhaps most strikingly in the case of \HNLs and a light scalar field, if one takes the simplest\footnote{An admittedly subjective statement.} model of neutrino masses, the type-I seesaw mechanism \cite{Gell-Mann:1979vob,Yanagida:1979as,deGouvea:2016qpx}, then we would generically expect a Yukawa coupling between \HNLs and the scalar field $\mathcal{L} \supset g \phi N^c N^c$, with $N^c$ the \HNL field written in two-component notation. Radiative corrections to the scalar mass, $m_\phi$, are now proportional to $g \mN$, with $\mN$ the mass of the fermions, rather than the soft scale $A$. Since it is essential that $\phi$ remains light so as to be a viable dark matter candidate, and $\mN$ may be large, this posses a problem for dark matter phenomenology. The problem is further worsened by the fact that the zero-temperature vacuum expectation value (vev) of the scalar field $\varphi_0= \langle \phi \rangle_{T=0}$ is large, 
\begin{equation}    
    \label{scalar-vev}
	\varphi_0 = -\frac{A {\rm v}^2}{m_\phi^2} ~\qq{where} {\rm v}^2 = \langle |H|^2 \rangle ~. 
\end{equation}
Thus, even for very small couplings i.e.\ $g\ll 1$, if $m_\phi$ is light then the vev of the scalar will induce a sizeable Majorana mass for $\mN$. This feature is not specific to type-I seesaw models and is generic to any mechanism that generates massive Majorana active neutrinos. Dirac neutrinos would conserve lepton number and forbid a coupling between the right-handed partners and $\phi$, and therefore represent a neutrino mass mechanism which is ``safe'' insofar as the scalar dynamics are concerned.\footnote{One could consider a charged Majoron, however this would necessarily carry two units of lepton number and be forbidden from coupling to the Higgs via the super-renormalizable portal.}  

The above discussion suggests that $m_\phi$ will depend on the scalar vev, $\varphi_0$, but \cref{scalar-vev} then tells us that the scalar vev must be determined self consistently. Indeed, assuming the vev induced mass of the \HNLs dominates, we may take $m_{N}\sim g \varphi_0$ (with, again, a subscript zero denoting zero temperature), such that the quantum correction to the scalar mass would be given by,
\begin{equation}
	m_\phi^2 \sim \frac{6g^4}{16\pi^2}  \varphi_0^2~\times(\log)~.
\end{equation}
If we again assume this radiative correction dominates over the bare mass, substitution into \cref{scalar-vev} then allows one to self-consistently estimate $\varphi_0$ in terms of $g$ and $A$. Radiative corrections therefore supply a mechanism by which the scalar potential self-adjusts towards a stable solution. If the scalar is too light, its vev will become large. This will induce a larger \HNL mass, which in turn radiatively generates a larger scalar mass. This then reduces the scalar vev to a smaller value.

\begin{figure}
    \centering
    \includegraphics[width=0.7\linewidth]{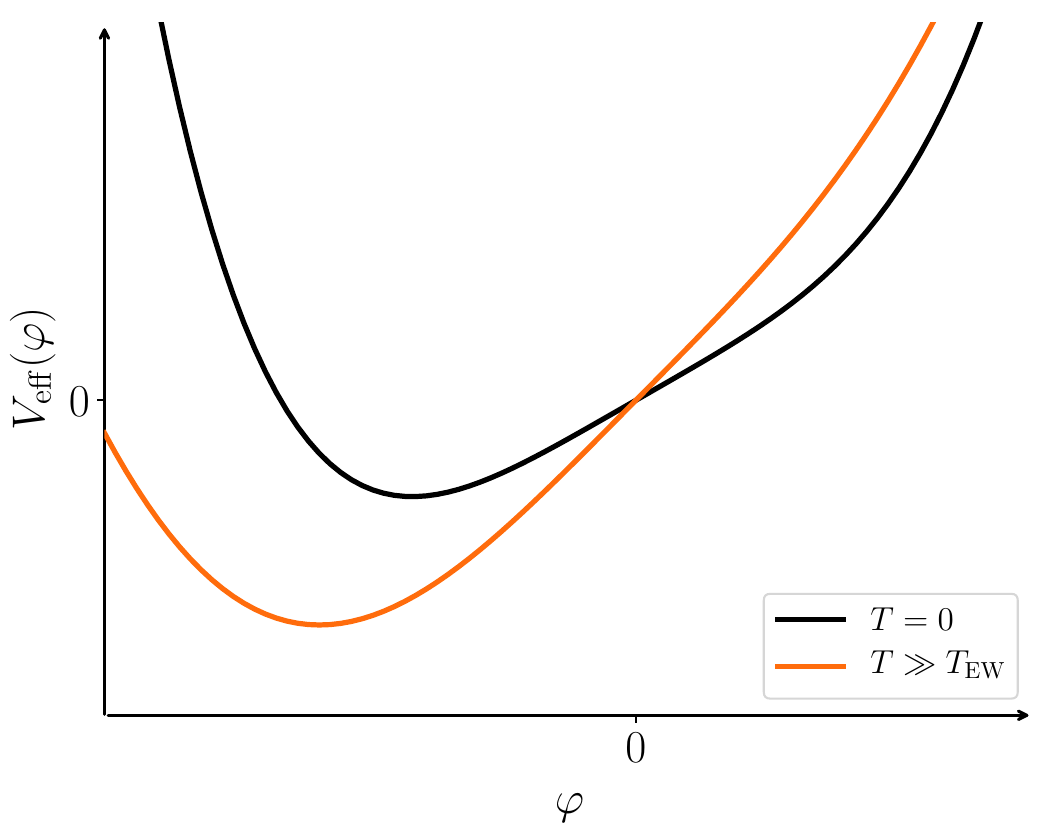}
    \caption{The effective potential, $V_{\rm eff}(\varphi)$, at zero-temperature (solid black line), and at finite temperature (solid orange line) in the vicinity of the minimum. We use $\varphi$ for the homogeneous mode of the scalar field. } 
    \label{fig:CWsketch}
\end{figure}

The above is simply a clumsy discussion of symmetry breaking in the presence of radiative corrections. As has been long understood, a systematic and efficient treatment of this effect is provided by the Coleman-Weinberg (CW) potential \cite{Coleman:1973jx}. Indeed, a re-interpretation of the above discussion may simply be rephrased as the {\it stabilization} of the scalar vev via radiative corrections. This is slightly different than the typical radiatively induced symmetry breaking that one associates with the CW model. A qualitative picture of the relevant effective potential is shown in \cref{fig:CWsketch} both at zero and finite temperature. 

In this paper we focus on the limit in which the \HNL and scalar mass are both dominated by radiative corrections. This is equivalent to considering the $O(1)$ ``patch'' of parameter space in which the bare fermion and scalar masses are sufficiently small, but nevertheless non-zero.  This has the advantage of being highly predictive. Ignoring matrix structure, two parameters control the dark matter abundance. We choose to work in terms of the Higgs-portal coupling $A$, and the physical scalar mass $m_\phi$ at $T=0$.  Fixing the dark matter relic abundance determines $A$ as a function of $m_\phi$ (assuming initial conditions are erased dynamically). Neutrino masses are generated via a standard seesaw mechanism, and the relevant neutrino textures accommodated by appropriately choosing the Yukawa couplings to charged leptons, $\mathcal{L} \supset Y L \tilde H N^c$. An investigation of the broader parameter space of the model  lies beyond the scope of this paper, and will be pursued separately. 

\subsection{Summary of results} 
Before entering the details of the paper, let us sketch our main conclusions. Focusing on regions of parameter space where the scalar is light, {and demanding small corrections to the mass from right-handed neutrinos,\!\footnote{See \cref{footnote1}.} pushes us towards the weak coupling limit in which $g\ll 1$.  We take $A$ and $g$ as two independent parameters which together radiatively generate the rest of the terms in the Lagrangian. This model is attractive because: {\it i)} it is radiatively stable below the electroweak scale, {\it ii)} its capable of explaining the dark matter relic abundance and neutrino masses, and {\it iii)} these explanations can be linked via the model's radiative (i.e., Coleman-Weinberg) dynamics.

For a fixed $A$ and  small coupling, $g \ll 1$, the radiative stabilization mechanism described above dynamically generates the following hierarchy of scales
\begin{equation}
    \label{hierarchy} 
	\underbrace{\varphi_0}_{O(1)} \gg \underbrace{\mN}_{O(g)}  
	\gg \underbrace{m_\phi}_{O(g^2)} ~, 
\end{equation}
where we measure scales relative to the vev of the field.  This hierarchy occurs because the \HNL mass arises at tree level, and the scalar mass at one loop from the fermion mass. Noting that $\varphi_0\propto A \vEW^2/m_\phi^2$ then fixes the formal scaling of $A \vEW^2$ with respect to $g$ since we have counted $\varphi_0$ as $O(1)$. This scaling demands $A \ll m_\phi$ such that the Higgs-portal contribution to the scalar mass is negligible relative to the \HNL mediated loop. It is also interesting to interpret \cref{hierarchy} with $A$ viewed as an input. From this IR perspective $A$ parameterizes a soft deformation of the secluded theory (with no coupling to the SM) which subsequently induces a tower of increasingly UV scales, each parametrically larger by factors of $1/g^n$. It is the interplay between this soft breaking, the small Yukawa coupling $g$, and the electroweak scale which induces the hierarchy outlined above. 

Remarkably, we find that the radiatively generated parameter space of the model is able to accommodate the observed relic abundance of dark matter in regions of parameter space that are currently untested. We find that the relic energy density in the coherent oscillating $k=0$ mode of $\phi$ is given by (up to a slowly varying logarithm)
\begin{equation}
    \Omega_\phi h^2 = 0.2 \bigg(\frac{A}{1~\mu{\rm eV}}\bigg)^2  \bigg(\frac{m_\phi}{3~{\rm meV}}\bigg)^{-11/4}  ~. 
\end{equation}
The parameter space differs substantially from the predictions of \cite{Batell:2022qvr} because the dark matter evolves via a non-linear differential equation due to the quartic (and higher order) terms in the CW potential. This introduces a number of other interesting dynamical features which we discuss below. 

\begin{figure}[h]
    \centering
    \includegraphics[width=0.8\linewidth]{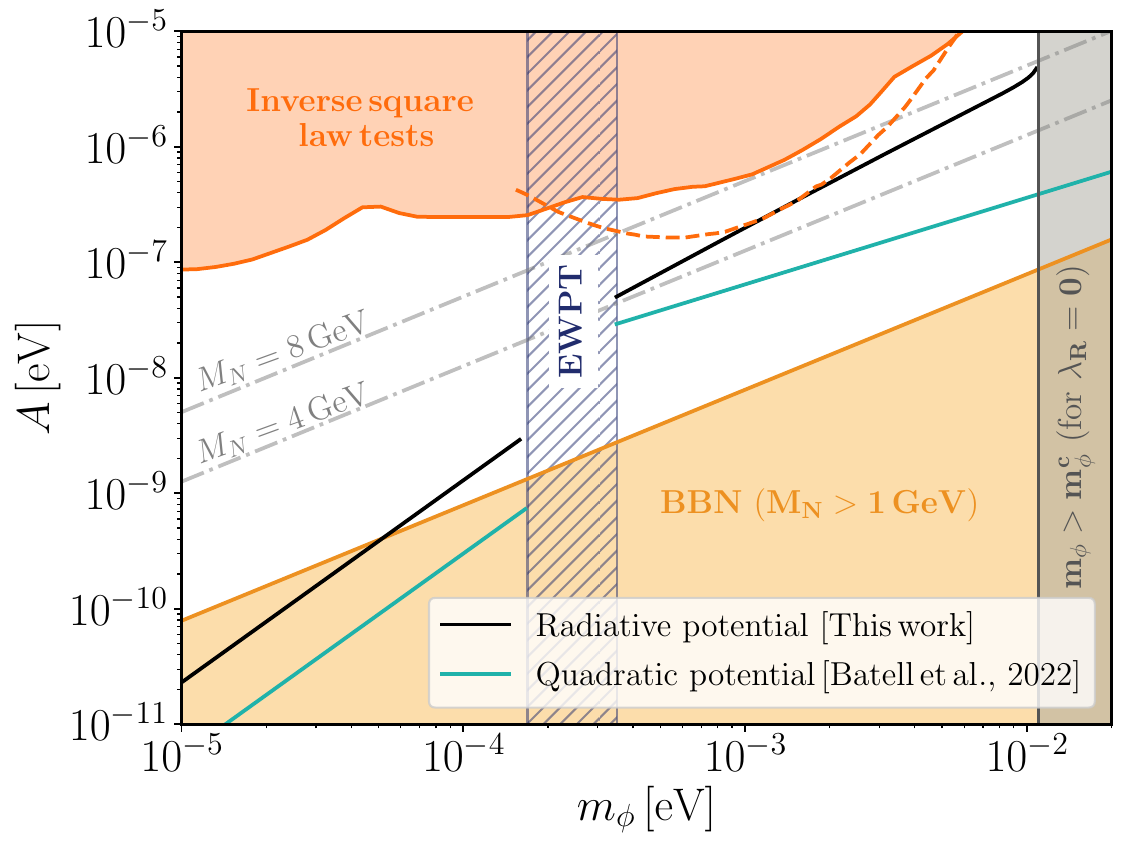}
    \caption{Parameter space in the $m_\phi$-A plane. The solid, black line corresponds to a relic density of $\Omega_\phi h^2 = 0.12$~\cite{Planck:2018vyg} derived using \cref{eq:RD_1,eq:RD_2}. For comparison, we show the results in a quadratic potential (teal), taken from \cite{Batell:2022qvr}. BBN constraints demand $(\Gamma_N)^{-1} < 0.1~ {\rm s}$~\cite{Ruchayskiy:2012si} which we impose by demanding $\mN \geq 1~{\rm GeV}$. Dynamics near the electroweak phase transition do not admit a simple analytic description ($100 \text{GeV} < T_{\rm osc} < 170$ GeV). In orange we show inverse square law tests  (solid orange) \cite{PhysRevLett.124.051301,PhysRevLett.98.021101,PhysRevLett.124.101101,PhysRevD.32.3084} and a future projection (dashed orange) from the HUST group \cite{Du:2022veu}. 
    \label{fig:plane}}
\end{figure}

Finally, perhaps most interesting from the perspective of laboratory tests of the model, we find that $\mN$ {\it always lies below the weak scale}. More quantitatively, when expressed in terms of the zero temperature mass, $m_\phi$, and the Higgs-portal coupling $A$, we find 
\begin{equation}
    \mN \sim \qty(\frac{A}{m_\phi})^{1/2} \vEW~.
\end{equation}
Benchmarking against the predictions of $A(m_\phi)$ from the dark matter relic abundance, we generically predict \HNLs in the mass range $\mN \in [1~{\rm GeV} , 10~{\rm GeV}]$. 

\subsection{Outline of paper}
The rest of the paper is organized as follows: 
In \cref{Model} we present the Lagrangian, and discuss the model's zero temperature dynamics.  Next, in \cref{DM-Dynamics} we consider the dynamics of the scalar field in the early universe. The bulk of our analysis is contained in \cref{Abundance,VEV-Mis} where we predict the dark matter relic abundance. In \cref{Pheno} we study experimental tests of the model we consider here, including fifth force and \HNL searches. Finally, in \cref{Conclusions} we summarize our findings and comment on future directions.

\section{Model definition and dynamics at zero temperature \label{Model}} 
The model we consider in this work is fully specified by the following Lagrangian,
\begin{equation}
    \label{model-def}
    \begin{split}
    \mathcal{L} = \frac12 (\partial \phi)^2 -V(\phi)+ \overline{N}^c\sigma_\mu \partial^\mu N^c &- \tfrac12 M_{ij} N^c_i N^c_j- \tfrac12g_{ij}\phi N^c_{i} N^c_{j}\\
    &-Y_{ij} \tilde{H} L N^c  
    - A \phi |H|^2 - \tfrac12 B \phi^2 |H|^2 ~+~{\rm h.c.} , 
    \end{split}
\end{equation}
where $V(\phi) = c_2 \phi^2 + c_3 \phi^3 + c_4\phi^4$;  we do not consider higher dimensional operators in this analysis.\footnote{Strictly speaking quantum corrections destabilize the effective potential at large field values, and higher dimensional operators may be required to ensure a bounded Hamiltonian. The dynamics we consider here are insensitive to these choices, and we do not discuss them further.} The scalar field $\phi$ is our dark matter candidate, and the \HNLs generate neutrino masses via the Type-I seesaw mechanism.

We are interested in the parameter space of this model in which $M_N$ is generated via a vev of $\phi$, and $m_\phi$ arises radiatively. It is therefore convenient to impose a discrete symmetry on the model which does not forbid the scalar \HNL coupling, and is softly broken by the super-renormalizable Higgs portal coupling $A$. For example, consider a $\mathbb{Z}_4$ symmetry where $\phi$ carries charge $2$ and $N^c$ carries charge $1$. If we identify this as a discrete lepton number, and charge the SM leptons appropriately, then the Yukawa interactions $YH LN^c$ are also allowed. We then have
\begin{align}
    \label{model-def-Z4}
    \begin{split}
    \mathcal{L}_{\mathbb{Z}_4} &= \tfrac12 (\partial \phi)^2 -c_2 \phi^2 -c_4 \phi^4 + \overline{N}^c\sigma_\mu \partial^\mu N^c \\
    & \hspace{2cm} - \tfrac12g_{ij}\phi N^c_{i} N^c_{j}-Y_{ij} \tilde{H} L N^c   + B \phi^2 |H|^2+~{\rm h.c.}, 
    \end{split}
\end{align}
We may then softly break the $\mathbb{Z}_4$ symmetry with 
\begin{equation}
    \mathcal{L}_{\rm soft}= - A \phi |H|^2 ~,
\end{equation}
which guarantees all symmetry breaking terms are proportional to powers of $A$. We estimate the size of the various radiatively induced parameters in \cref{app:rad_corr_est}; provided the bare parameters in \cref{model-def} are smaller than the estimates presented in \cref{app:rad_corr_est} we expect the model to be stable against radiative corrections. As has been discussed previously in the literature \cite{Piazza:2010ye}, in order for the renormalizable Higgs portal to not induce a large radiative mass for $\phi$, we require $B \ll 4\pi m_\phi/m_h$. A typical radiatively induced value for this coupling is $B\sim Y^2 g^2/16\pi^2$, which easily satisfies the constraint of the previous sentence in the parameter space of interest to us. For the Coleman-Weinberg dynamics we discuss to be operational, one should have $c_4 \lesssim  g^4/16\pi^2$. 

\subsection{Coleman-Weinberg potential}
Let us now analyze the CW potential of the dark sector in the secluded limit. We will always use $\varphi$ when referring to the classical dynamics of the scalar field, and $\phi$ to denote quantum excitations. Following the standard construction i.e.\ introducing background field dependent masses, and appropriate counterterms, we then fix the Coleman-Weinberg potential with the following renormalization conditions (with primes denoting derivatives) \cite{Schwartz:2014sze}
\begin{align}
    \qty[ V_{\rm CW} ]_{\varphi=0} &=\Lambda^4~, \label{ren-1}\\
    \qty[ V''_{\rm CW} ]_{\varphi=0} &=0~, 
    \label{ren-2}\\
    \qty[ V''''_{\rm CW} ]_{\varphi=\varphi_R} &=\lambda_R ~
    \label{ren-3}.
\end{align}
\cref{ren-2} {\it defines} the boundary between a broken and unbroken phase in the classical theory. The pole mass, $m_\phi$, is defined as the curvature of the effective potential at {\it its minimum} i.e.\ not at $\varphi=0$. At the order we are working the renormalization conditions for $g$ and $M$ ({\it cf}. \cref{model-def}) are immaterial. Setting $M$ to zero, we arrive at the renormalized CW potential
\begin{equation}
    V_{\rm CW}(\varphi; A=0)= \frac{1}{4!} \varphi^4 \left(\frac{75 g^4}{8\pi ^2}+\lambda_R\right)-\frac{3g^4\varphi^4}{32 \pi^2}\log\qty(\frac{\varphi^2}{\varphi_R^2}) + \Lambda^4 ~.
\end{equation}
The \HNLs are fermions, and their quantum corrections do not induce symmetry breaking. Let us note that the stability of the theory depends on the choice of $\lambda_R$ and $\varphi_R$, which is unsurprising; a negative quartic coupling will have a potential that is unbounded from below. In fact, these two quantities together form a renormalization group (RG) invariant quantity. If $\varphi_R$ is varied, then $\lambda_R$ adjusts itself such that physical predictions are unaffected. 

Let us now softly break our $\mathbb{Z}_4$ symmetry with the introduction of the Higgs-portal coupling proportional to $A$. The dominant contribution to the cubic coefficient comes from a loop of \HNLs and is given by $c_3 \sim \frac{1}{16\pi^2} g^3 m_N = \frac{1}{16\pi^2} g^4 \varphi_0$. One can check {\it a posteori} that for $\varphi_0$ given by \cref{vev-g} this is very small.   
Therefore, even in the presence of the Higgs-portal, neglecting the cubic interaction of the scalar potential is still an excellent approximation. 

The parameters we consider here are such that the Higgs field's dynamics are only slightly perturbed by the presence of the secluded sector. We may therefore set the Higgs field to its Standard Model vev at $T=0$. Since $A/m_H\ll 1$ it is legitimate to neglect loop corrections from the Higgs portal. The CW potential then assumes the form,
\begin{equation}
    \label{Full-CW}
    V_{\rm CW}(\varphi)= A |H|^2 \varphi + \frac{1}{4!} \varphi^4 \left(\frac{75 g^4}{8\pi ^2}+\lambda_R\right)-\frac{3g^4\varphi^4}{32 \pi^2}\log\qty(\frac{\varphi^2}{\varphi_R^2}) +\Lambda^4~.
\end{equation}
Higher order corrections proportional to $\lambda_R^2$ would appear in a more general treatment, however, we take $\lambda_R^2\sim O(g^8)$ in our counting (i.e.\ assuming $\lambda_R$ to be radiatively generated) and so we neglect these contributions. Depending on the choice of $\lambda_R$ the potential either has a minimum at $\varphi_0 < 0$ or a maximum at $\varphi_0 > 0$. In both cases $|\varphi_0|\sim O(g^{-4/3})$. We demand that $\lambda_R$ be such that the potential has a minimum at a negative value of $\varphi_0$. It is convenient to choose $\varphi_R\sim (A \vEW^2/g^4)^{1/3}$, such that $\log(\varphi_0/\varphi_R)$ is $O(1)$ in the vicinity of the minimum, in which case requiring $\lambda_R \geq -\frac{75 g^4}{8\pi ^2}$ is sufficient. The potential minimum in our model will generally be meta stable, but with an extremely high potential barrier. 

\subsection{Mass spectrum}
In what follows, we make the RG-invariant choice: $\lambda_R=0$ and $\varphi_R=- \sqrt[3]{(8 \pi ^2 A \vEW^2)/(11 g^4)}$. These parameter choices do not qualitatively affect the discussion in what follows beyond ensuring the potential has a stable minimum.\footnote{To remain in a ``radiatively generated'' region of parameter space the scaling $\lambda_R\sim O(g^4)$ should be respected.}  The vev of $\phi$ is then given by 
\begin{equation}
    \varphi_0 = \varphi_R  = -\qty(\frac{8\pi^2A \vEW^2}{11 g^4})^{1/3}~, \label{vev-g}
\end{equation}
while the mass of the scalar at $T=0$ (i.e.\ the quantum fluctuations around the minimum) is given by 
\begin{equation}
    m_\phi^2 = \qty[V''(\varphi)]_{\varphi=\varphi_0} = \frac{27}{2}\qty(\frac{A \vEW^2 g^2}{11\pi})^{2/3} ~.
\end{equation}
We may then trade $g$ for $m_\phi$, 
\begin{equation}
    g= \frac{2^{3/4}(11 \pi)^{1/2} }{ 3^{9/4} } \qty(\frac{m_\phi^3}{A\vEW^2})^{1/2} .
\end{equation}
Above quantities can then be expressed in terms of the pole mass rather than the Yukawa $g$, for example \cref{vev-g} may be re-written as 
\begin{equation}
    \label{vev-m}
    \varphi_0 = \varphi_R  = -\frac{27}{11} \frac{A \vEW^2}{m_\phi^2} ~. 
\end{equation}
Notice that this differs by an $O(1)$ factor from the relationship between $A$, $m_\phi$, and $\varphi_0$ when $m_\phi$ is dominated by its tree-level value \cite{Piazza:2010ye}. 

Finally, the physical mass of the \HNLs (again neglecting matrix structure) is given by  
\begin{equation}
    \label{RHN-mass}
    \mN= g|\varphi_0| = 6^{3/4}\sqrt{\frac{\pi}{11}} \qty(\frac{A}{m_\phi})^{1/2} ~\vEW~.
\end{equation}
This is our first major result: the mass of the \HNLs is parametrically tied to the weak scale. The ratio $A/m_\phi \ll 1$ when the mass of $\phi$ is radiatively generated. We therefore conclude that the mass of \HNLs will always lie below the weak scale and can therefore be searched for using accelerator and fixed target facilities (see \cref{RHN-search}). 

Neutrino masses are generated via the standard seesaw mechanism. Assuming no special flavor structure in the \HNL mass matrix we expect the standard type-I seesaw formula to apply 
\begin{equation}
    m_\nu^{ij} = m_D^{i \alpha} M^{-1}_{\alpha\beta} m_D^{\beta j}~,
\end{equation}
where $m_D = Y \vEW$ is the Dirac mass matrix connecting active and sterile neutrinos, and all indices run from 1 to 3;  the raising and lowering of indices is done to de-clutter the notation. As we discuss below, the Yukawa textures play no substantial role in the dark matter dynamics. For instance, a pseudo-Dirac pair of \HNLs may allow for an inverse seesaw mechanism to operate, such that $m_\nu \sim m_D^2 M^{-2} \mu$ where $\mu$ is the mass splitting between the pseudo-Dirac pair~\cite{Mohapatra:1986aw,Mohapatra:1986bd}. This scenario yields a small mass splitting which can be advantageous for leptogenesis~\cite{Akhmedov:1998qx,Klaric:2020phc}. 

As an aside, let us also note that the model has an additional radiative neutrino mass mechanism. A diagram with the Higgs running in the loop with a single vev insertion of $\phi$ yields a neutrino mass that scales as 
\begin{equation}
    m_\nu \sim \frac{6 g Y^2}{16\pi^2} \varphi_0 \log(m_\nu^2/m_H^2)~, 
\end{equation}
Although we generically expect this to be sub-dominant to the standard seesaw mass, it may contribute appreciably when $A/m_\phi$ is not too small, or when there is fine tuning such that $|g|~ Y_{ij}Y_{ji} \ll Y_{ij} g_{jk} Y_{ki}$ with $|g|$ the typical value of an entry in  $g_{jk}$. For the parameter space preferred by the dark relic abundance, this contribution is negligible and we do not consider it further.

\subsection{Decay rates}
Let us now turn to the properties of the \HNLs. Because of their couplings to the scalar field, a new decay pathway is available. For two of the \HNLs we will have  $N_i\rightarrow N_j \nu$ with $j\leq i$ (assuming labeling is ordered in mass). For the lightest \HNL the decay pathway $N\rightarrow\nu \phi$ is available and scales as 
\begin{equation}
    \Gamma_{N\rightarrow \nu \phi} = \frac{g^2 \mN}{8\pi}  \sum_{\alpha} \theta_\alpha^2  ~,
\end{equation}
where $\theta_{\alpha}$ is the mixing between the state $N_1$ and the three active neutrinos. This rate should be compared against the muon-like decay formula 
\begin{equation}
    \Gamma_{N\rightarrow \ell \ell \nu} = \frac{G_F^2\mN^5}{192 \pi^2} \sum_{\alpha} \theta_\alpha^2~.
\end{equation}
The same mixing angle appears in both cases, and so we see the relevant comparison is $g^2/8\pi$ vs $G_F^2 \mN^4/(192\pi^2)$. In what follows we find that $ g^2 \ll G_F^2 \mN^4$ such that the \HNL's lifetime is not substantially modified by the scalar decay mode. 

The scalar is itself unstable. The decay rate is given parametrically by $\Gamma_\phi \sim \theta^4 g^2 m_\phi$. For 
a scalar mass around $1~{\rm meV}$ and assuming a single massless active neutrino such that $\phi \rightarrow \nu \nu$ is allowed,\footnote{If all neutrinos are heavier than $\phi$, then $\phi \rightarrow \gamma \gamma$ is the leading decay channel. This is even slower than $\phi \rightarrow \nu \nu$ and never threatens dark matter stability on cosmological timescales~\cite{Piazza:2010ye}.} the scalar lifetime is given roughly by
\begin{equation}
    \tau_\phi \sim 3 \times 10^{19} ~{\rm Gyr} \left(\frac{5 \times 10^{-13}}{g}\right)^2 \left(\frac{1 ~{\rm  meV}}{m_\phi}\right) \left(\frac{10^{-5}}{\theta}\right)^4~.
\end{equation}
Comparing to the age of the universe, $\tau_{U} \sim 13.8 ~{\rm Gyr}$, the dark matter is clearly stable on cosmological timescales.

\subsection{Scalar couplings to matter} 
Finally, let us briefly review constraints on a light scalar mixing with the Higgs. More detailed discussions can be found in \cite{Piazza:2010ye,Batell:2022qvr}. The $A \phi |H|^2$ coupling induces mixing between the Higgs boson, $h$, and $\phi$ after electroweak symmetry breaking. The coupling of the Higgs to matter then induces couplings between $\phi$ and matter.

The largest such coupling for nucleons comes from heavy quarks, which is transmitted to hadrons via an anomaly matching condition \cite{Shifman:1978zn,Cheng:1988cz,Barbieri:1988ct}. In the low energy description the coupling is dominantly to gluons. Electron couplings are simply proportional to the electron Yukawa. The result is that $\phi$ couples to nucleons and electrons with a strength given by 
\begin{equation}    
    \label{nucleon-scalar}
    g_{\phi N N} \sim \frac{A \Lambda_{\rm had} }{m_h^2}~,\quad g_{\phi ee} = \frac{A m_e}{m_h^2}~,
\end{equation}
where $\Lambda_{\rm had}$ is a typical hadronic scale on the order of the nucelon mass. These couplings can induce a Yukawa-like fifth force between test bodies and be used to search for an ultralight scalar \cite{PhysRevLett.124.051301,PhysRevLett.98.021101,PhysRevLett.124.101101,PhysRevD.32.3084}. 
\section{Dynamics in the early universe \label{DM-Dynamics} } %
In the original proposal of Piazza and Pospelov \cite{Piazza:2010ye} the dark matter abundance was set by a choice of initial conditions for the misalignment of the scalar field. Oscillations onset at $3H=m_\phi$ and the dark matter relic density can be estimated straightforwardly given the initial misalignment. Batell, Ghalsasi, and Rai  demonstrated in \cite{Batell:2022qvr} that there exists a regime in which the misalignment that determines the relic abundance is dictated by thermal properties. This then supplies a predictive relic density target, at least so long as initial misalignment is not too large (see \cref{init-cond-footy}). We also note that the ``freeze-in'' population of scalars $\phi$ is highly suppressed. This can be easily seen by comparing the dominant production channel $q\bar{q} \rightarrow H \rightarrow H \phi$ through the Higgs portal $\langle n \sigma v \rangle \sim |A|^2/T$ against Hubble rate $H\sim T^2/M_{\rm Pl}$.

The dynamics of the scalar field in the CW potential differ qualitatively from the quadratic case considered in \cite{Piazza:2010ye,Batell:2022qvr}. In what follows we describe the evolution of the scalar field. We focus on initial conditions that are sufficiently small such that thermal misalignment dominates (see \cref{Preheating}). Much of our analysis mirrors the setup in \cite{Batell:2022qvr} and so we do not belabor points that are discussed in detail there. What is new is the shape of the effective potential, being stabilized by radiative corrections rather than by the bare scalar mass, and new degrees of freedom in the form of \HNLs.

\subsection{Thermal misalignment \& relic abundance \label{Abundance} } 
In this section we consider the relic abundance generated by the thermal misalignment mechanism assuming that initial conditions are sufficiently small such that the dark matter misalignment is dominated by thermal effects. We focus on parameter space where oscillations begin before the electroweak phase transition.  We consider the evolution of the field's $k=0$ mode, $\varphi:= \phi_{k=0}$ using the equations of motion, 
\begin{equation}
    \ddot{\varphi} + 3 H \dot{\varphi} +  \pdv{\varphi} V_{\rm eff}(\varphi)=0 ~,
\end{equation}
where dotted derivatives correspond to $\dv{t}$. It is convenient to recast this equation in terms of temperature in a radiation dominated epoch, where $H= \gamma T^2/M_{\rm Pl}$ where we use the reduced Planck mass $M_{\rm Pl}=2.43 \times 10^{18}~{\rm GeV}$ and $\gamma= \sqrt{\pi^2 g_*(T)/90}$. Using the Jacobian $\dd T/\dd t  = - HT$, we can rewrite the time derivatives as temperature derivatives,\footnote{Using  %
\[
    \dot{\varphi}  =  \dv{\varphi}{T}\dv{T}{t} = - \dv{\varphi}{T} H T, \qq{and}
    \ddot{\varphi}  = \dv{t} \left(-\dv{\varphi}{T} H T\right) = \dv[2]{\varphi}{T} H^2 T^2 + 3 \dv{\varphi}{T} H^2 T.
\].}  which leads to 
\begin{equation}
    \dv[2]{\varphi}{T} = - \frac{1}{H^2T^2} \frac{\partial V_{\rm eff}}{\partial \varphi}~ = -\frac{M_{\rm Pl}^2}{\gamma^2 T^6}~\frac{\partial V_{\rm eff}}{\partial \varphi}~.
    \label{eq:eom1}
\end{equation}
In practice we set $g_* \approx 106.75={(\rm const)}$ for numerical estimates above the electroweak scale~\cite{Borsanyi:2016ksw}. The effective potential contains contributions from the zero temperature  and finite temperature potentials. For the finite temperature piece we make use of the thermal functions $J_B$ and $J_F$ for bosons and fermions, respectively~\cite{Quiros:1999jp}. At early epochs the only thermal degree of freedom with coupling to $\phi$ is the Higgs and so, following \cite{Piazza:2010ye}, we include only the contributions of the Higgs field and the Goldstone bosons, $\chi$, in Landau gauge
\begin{equation}
    V_{\rm eff} = V_{\rm CW}(\varphi) + \frac{1}{2\pi^2} \, T^4 J_B\left[\frac{m_h^2 (\varphi,h,T)}{T^2}\right] + \frac{3}{2\pi^2} \, T^4 J_B\left[ \frac{m_\chi^2(\varphi,h,T)}{T^2}\right]~.
\end{equation}
The \HNLs can also contribute if they thermalize, but this turns out to happen after the electroweak phase transition; we defer a discussion to \cref{Thermal-HNLs}. The effective masses are given simply by $m^2=m_0^2 +\Pi(T)$, where $m_0^2$ is the zero temperature effective mass and $\Pi(T)$ is the thermal self-energy. The only dependence on $\varphi$ enters in the zero temperature effective mass,
\begin{align}
    m_{h,0}^2(\varphi,h)&= -\mu^2 + 3\lambda h^2 + A \varphi~,\\
    m_{\chi,0}^2(\varphi,h)&= -\mu^2 + \lambda h^2 + A \varphi~,
\end{align}
where $h$ is the value of the Higgs field's $k=0$ mode. The self-energies are proportional to $T^2$,
\begin{equation}
\Pi(T) = T^2\left(\frac{3}{16}g^2 + \frac{1}{16}g'^2 + \frac{1}{4}y_t^2 + \frac{1}{2}\lambda^2 \right) ,    
\end{equation}
where $g$, $g'$, $y_t$ and $\lambda$ are the electroweak gauge coupling, the hypercharge gauge coupling, the Yukawa coupling of the top quark and the quartic coupling of the standard model Higgs potential, respectively.
At high temperatures $\Pi(T)$ dominates the argument of the bosonic thermal function, and $J_B'[\Pi(T)^2/T^2] \approx 0.5$ (numerically)~\cite{Batell:2022qvr}. The derivative of the effective potential is then given by  
\begin{equation}
\begin{split}
    \frac{\partial V_{\rm eff}}{\partial \varphi} & =\frac{1}{8\pi^2}g^4 (11-3{\sf L}) \varphi^3 + \frac{A}{2\pi^2}T^2\left(J_B' \left[\frac{m_h^2(\varphi,h,T)}{T^2}\right] + 3 J_B'\left[\frac{m_\chi^2(\varphi,h,T)}{T^2}\right]\right) \\
    & \simeq \frac{1}{8\pi^2}g^4 (11-3{\sf L}) \varphi^3 + \frac{A}{\pi^2}T^2+ \ldots ~ \qq{for} T\gg \vEW~~, 
\end{split}
\end{equation}
where ${\sf L}=\log(\varphi^2/\varphi_R^2)$. This is the thermal misalignment mechanism introduced in \cite{Buchmuller:2004xr,Batell:2021ofv}. The linear piece tilts the potential at the origin, resulting in a temperature dependent minimum 
\begin{equation}
    \varphi_{\rm min}(T) \simeq -\frac{27}{(11\pi)^{2/3}} \left( \frac{1}{11-3{\sf L}_{\rm min}}\right)^{1/3}\frac{A\vEW^2}{m_\phi^2}\left(\frac{T}{\vEW}\right)^{2/3}~,
\end{equation}
where ${\sf L}_{\rm min} = \log(\varphi_{\rm min} ^2/\varphi_R^2)$ can be determined self consistently via an iterative procedure. The thermally induced tilt drives the field to negative values even in the presence of Hubble friction and provides a mechanism for erasing initial conditions.\footnote{ \label{init-cond-footy} Thermal misalignment can generically erase initial conditions on the order of the zero temperature vev $\varphi_0 \sim A \vEW^2/m_\phi^2$ which is typically on the order of $10^{13}~{\rm GeV}$ for the parameters we consider.}

At high temperatures, Hubble friction is large and the field's dynamics are dominated by the thermal misalignment. The initial phase of thermal misalignment is then given by,
\begin{equation}\label{initial-phi}
    \varphi_{\rm{pre}}(T)= -\frac{A M_{\rm Pl}^2}{6 \pi^2 \gamma^2 T^2}  + \varphi_{I}~ \qq{for} T\gg T_{\rm osc} ~,
\end{equation}
where $T_{\rm osc}$ is the temperature at which $\varphi$ starts oscillating, and will be defined below. The initial condition,  $\varphi_{\rm I}$, is set  at some large temperature where $\dot{\varphi}_I=0$. As $T$ decreases, the field value drifts towards increasingly negative values. 

The solution in \cref{initial-phi} is valid at high temperatures while the Hubble friction is still effective. Examining the equations of motion,\footnote{In a harmonic potential one would have $\ddot{\varphi} + 3H \dot{\varphi} + m^2 \varphi = F $,  with $F$ an ``external force''. Whereas we have  $\ddot{\varphi} + 3H \dot{\varphi} + \partial_\varphi V_{\rm CW} \varphi = F ~.$ } one sees that $m^2_{\rm eff}(\varphi) = \partial_\varphi V_{\rm CW}(\varphi)/\varphi$ plays the same role as a fixed mass in a harmonic oscillator. The condition for the onset of oscillations is  $[3H]^2= [m_{\rm eff}(\varphi_{\rm pre})]^2$ evaluated at $T=T_{\rm osc}$,  where we use \cref{initial-phi} to compute $m_{\rm eff}^2$.  Using $m_{\rm eff}^2 = g^4\varphi_{\rm pre}^2(11-3{\sf L}_{\rm osc})/(8\pi^2)$ where ${\sf L}_{\rm osc} = \log(\varphi^2/\varphi_R^2)$ we find, 
\begin{equation}\label{eq:Tosc}
    T_{\rm osc} = \frac{{\rm v}}{3} \left( \frac{M_{\rm Pl} m_\phi}{\gamma {\rm v}^2}\right)^{3/4} \left(\frac{11^{1/4}  (11-3{\sf L}_{\rm osc})^{1/8}}{3^{5/8}~2^{1/4} \, \pi^{1/2}}\right)~,
\end{equation}
where ${\sf L}_{\rm osc}$ can be estimated via the same iterative procedure described above,
\begin{equation}
        {\sf L}_{\rm osc} \approx \log \left( \frac{11 m_\phi M_{\rm Pl}}{6 \pi^2 \sqrt{3} \gamma \sqrt{11-3\sf{L}_{\rm osc} } \vEW^2}\right)~.
\end{equation}
This iterative procedure converges provided $\varphi^2/\varphi_R^2\lesssim 39$ which is equivalent to demanding ${\sf L}_{\rm osc} \lesssim 11/3$. This can be re-written in terms of the scalar mass as $m_\phi \lesssim m_\phi^{(c)}$ with $m_\phi^{(c)} \sim 10~{\rm meV}$. The critical mass, $m_\phi^{(c)}$, depends on $\lambda_R$, and our numerics use $\lambda_R=0$. Above the critical mass, the tilt of the potential is too strong at the onset of oscillations and there is no local minimum. For $m_\phi \gtrsim 10~{\rm meV}$, the dynamics demand a tree-level quartic-coupling $\lambda_R>0$ to stabilize the potential at the relevant epoch. For $m_\phi \gg  m_\phi^{(c)}$, one would require $\lambda_R \gg O(g^4)$, such that the relevant parameter space is not radiatively generated from $A$ and $g$. 

Let us note that $\varphi_{\rm pre}(T_{\rm osc})$ is rather close the minimum $\varphi_{\rm min}(T_{\rm osc})$ such that a harmonic approximation is valid almost immediately after the onset of oscillations. The mass of $\phi$, which we define as $\mu^2_\phi(T)= [V''(\varphi)]_{\varphi=\varphi_{\rm min}}$, will be temperature dependent due to the drifting minimum 
\begin{equation}
    \mu_\phi^2(T)=   m_\phi^2 \left( 1 - \frac{{\sf L}_{\rm min} }{3}\right) \left(\frac{T}{\vEW}\right)^{4/3} \left(\frac{11/\pi^2}{11-3 {\sf L}_{\rm min} }\right)^{2/3}~.
\end{equation}
The slow variation of the minimum, $\varphi_{\rm min}\propto T^{2/3}$, relative to Hubble, $H\propto T^2$ and by proxy the oscillation frequency, means that the shifting of the potential minimum is effectively adiabatic. The scalar oscillates around the minimum and behaves like cold dark matter almost immediately. 

\begin{figure}[h]
    \centering
    \includegraphics[width=0.8\linewidth]{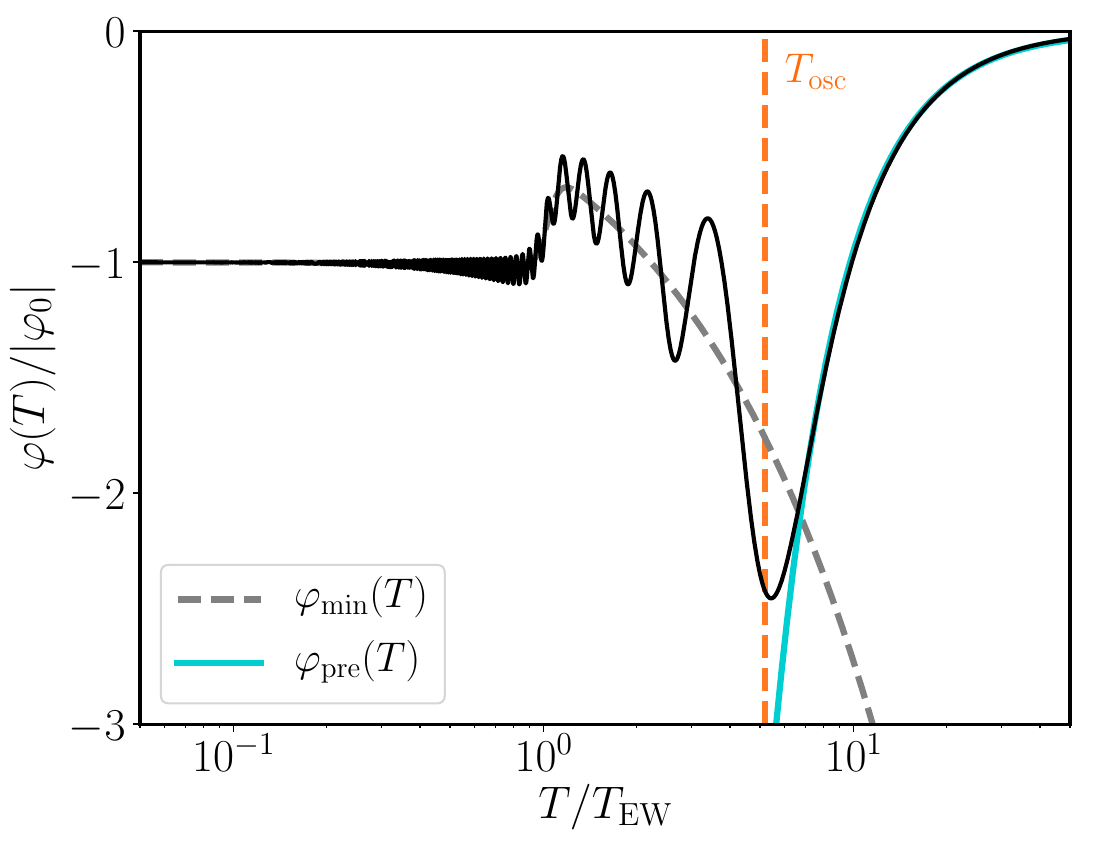}
    \caption{Dynamics of $\varphi(T)$ for  $m_\phi = 3$ meV (after re-scaling to $\varphi_0$ results are independent of $A$). The solid black line shows the numerical evolution of $\varphi$ with temperature. The gray dashed line shows the thermal evolution of the scalar minimum, and $\varphi_{\rm pre}$ is plotted in turquoise. We show the predicted onset of the oscillations in a vertical orange dashed line. The EW phase transition has been modeled by an hyperbolic tangent with width $15 \text{ GeV}$~\cite{DOnofrio:2015gop}. We take ${\sf L}=0$ for simplicity. }
    \label{fig:my_label}
\end{figure}

With this picture in mind it is clear that the amplitude of the scalar oscillations {\it relative to the minimum} determines the relic abundance. It turns out that this is given by 
\begin{equation}
    \varphi_{\rm osc} = | \varphi_{\rm pre}(T_{\rm osc})-\varphi_{\rm min}(T_{\rm osc})|=\qty(\sqrt[3]{3/2}-1)\times |\varphi_{\rm min}(T_{\rm osc})|\approx 0.145 \times |\varphi_{\rm min}(T_{\rm osc})| ~.
    \label{eq:relthermal}
\end{equation}
Numerical solutions of \cref{eq:eom1}  show that the inclusion of the temperature dependent Higgs vev, \vEW, simply adiabatically transfers the oscillating solution to the final zero temperature minimum as is shown in \cref{fig:my_label}. The relic scalar abundance $\Omega_\phi$ can be expressed in terms of model parameters.  The dynamics are adiabatic, such that number of particles per comoving volume is conserved and $n_\varphi /s ={\rm const}$ where $s=\tfrac{2\pi^2}{45}g_*(T) T^3$ \cite{Kolb:1990vq}. The number density at $T=T_{\rm osc}$ is given by $n(T_{\rm osc}) = \rho(T_{\rm osc}) / \mu(T_{\rm osc})$, where $\rho(T_{\rm osc}) \simeq \tfrac{1}{2} \mu_\phi^2(T_{\rm osc}) (\varphi_{\rm pre}(T_{\rm osc}) - \varphi_{\rm min}(T_{\rm osc}))^2$ is measured relative to the instantaneous minimum of the potential. The thermal misalignment therefore generates a dark matter abundance of
\cite{Preskill:1982cy,Kolb:1990vq,Mukhanov:2005sc}
\begin{equation}\label{eq:RD_1}
    \Omega_\phi h^2= \frac{\rho(T_{\rm osc})}{\rho_c} \qty(\frac{m_\phi}{\mu(T_{\rm osc}) }) \qty(\frac{T_{0}}{T_{\rm osc}})^3\frac{g_*(T_0)}{g_*(T_{\rm osc})}= 0.15 \bigg(\frac{A}{1~\mu{\rm eV}}\bigg)^2  \bigg(\frac{3~{\rm meV}}{m_\phi}\bigg)^{11/4} ~.
\end{equation}
Here, $\rho_c\approx 10^{-5}~{\rm GeV}/{\rm cm}^3$ \cite{Planck:2018vyg}, and $g_*(T_0)= 3.91$ while we take $g_*(T_{\rm osc}) =106.75$ \cite{Borsanyi:2016ksw}, and $T_0= 2.7 \text{ K}$ is the temperature of the universe today \cite{Fixsen_2009}.
For thermal misalignment to be operational, we require that $T_{\rm osc} \gtrsim T_{\rm EW}$. Otherwise the scalar field will be stuck by Hubble friction until after the electroweak phase transition. Demanding $T_{\rm osc} \geq 170~{\rm GeV}$ \cite{DOnofrio:2014rug,DOnofrio:2015gop} requires $m_\phi \geq 0.35~{\rm meV}$. We now turn to lower masses where oscillations begin after the electroweak phase transition. 


\subsection{VEV misalignment \label{VEV-Mis}}

Next consider the case where $T_{\rm osc} \ll  T_{\rm EW}$. In this limit the electroweak symmetry is broken, the Higgs' contribution to the thermal functions is Boltzmann suppressed, and the dynamics are governed by 
\begin{equation}
    \ddot \varphi + 3 H \dot \varphi + m_{\rm eff} \varphi = - A\vEW^2~.
\end{equation}
In the regime $T_{\rm osc} < T \ll T_{\rm EW}$, the zero-mode amplitude of the scalar field is given by\footnote{Initial conditions from $\varphi_{\rm pre}$ have a small effect for $T\ll T_{\rm EW}$ and so we omit them for simplicity.}
\begin{equation}
  \varphi_{\rm post}(T) \simeq - \frac{A M_{\rm Pl}^2 \vEW^2}{20 \gamma^2 T^4},  ~ \qq{for} T_{\rm osc}<  T \ll T_{\rm EW}~.
\end{equation}
The amplitude at which the field starts oscillating is given by the relative amplitude of the field at $T_{\rm osc}$ with respect to the minimum. To estimate $T_{\rm osc}$ in this regime we require $[m_{\rm eff}(\varphi_{\rm post})]^2 = (3H)^2$ at $T=T_{\rm osc}$, which leads to 
\begin{equation}
    T_{\rm osc} = \frac{11^{1/4}}{3^{11/12}20^{1/6}}\sqrt{\frac{m_\phi M_{\rm Pl}}{\gamma}}~.
\end{equation}
We note that the dependence on $m_\phi$ is different from $T_{\rm osc}$ above the electroweak scale (see \cref{eq:Tosc}). At the onset of oscillations, the energy density is $\rho (T_{\rm osc}) \simeq \tfrac{1}{2}m_\phi^2(\varphi_{\rm pre}(T_{\rm osc})-\varphi_0)^2$, where we have approximated the mass $\mu_\phi(T)$ and the minimum $\varphi_{\rm min}(T)$ of the scalar field by their zero temperature values. Note that the definition of $m_{\rm eff} = \partial_\varphi V_{\rm CW} / \varphi$ does not change with respect to the previous case. However, in this regime $[\partial_\varphi V_{\rm CW}]_{\varphi=\varphi_{\rm min}} \sim A \vEW^2$ and we recover $m_{\rm eff}(T \ll T_{\rm EW}) \sim O(1)  \times m_\phi$. Analogously to \cref{eq:relthermal}, 
\begin{equation}
  \varphi_{\rm osc} = |  \varphi(T_{\rm osc}) - \varphi_0| = \left( \frac{1}{5^{1/3}}\left(\frac{3}{2}\right)^{2/3}-1\right) \times |\varphi_0| \approx 0.234 \times |\varphi_0|~~, 
\end{equation}
which guarantees that oscillations begin, and remain, close to the minimum of the potential where a harmonic approximation is reliable. Knowing $\varphi_{\rm osc}$ we can estimate the post-electroweak contribution of the scalar oscillations to the relic density today,
\begin{equation} 
    \label{eq:RD_2}
    \Omega_\phi h^2 \simeq \frac{\rho(T_{\rm osc})}{\rho_c} \qty(\frac{T_{0}}{T_{\rm osc}})^3\frac{g_*(T_0)}{g_*(T_{\rm osc})} \simeq 0.1 \bigg(\frac{A}{1~{\rm neV}}\bigg)^2  \bigg(\frac{90~{\rm \mu eV}}{m_\phi}\bigg)^{7/2} ~,
\end{equation}
for $T_{\rm osc} \ll T_{\rm EW}$ and $g_*(T_{\rm osc}) \simeq 90$~\cite{Borsanyi:2016ksw}.\footnote{Within the parameter space plotted in \cref{fig:plane}, oscillations onset well before the QCD phase transition, and the relativistic degrees of freedom only change from $80 \lesssim g_* \lesssim 100$. We take $g_* \simeq 90$ for simplicity.} For $T\ll T_{\rm EW}$ the dynamics match closely with those in Ref.~\cite{Batell:2022qvr} for a quadratic potential. This is because the effective potential is well approximated by its zero temperature form, and the oscillations take place close to the minimum. Since we have defined $m_\phi^2$ as the curvature about the minimum, the dynamics are equivalent to a quadratic potential with a mass $m_\phi$ as studied in \cite{Batell:2022qvr}. In the above expression we have taken ${\sf L}_{\rm osc} \approx 0$ as $\varphi_{\rm post}(T_{\rm osc}) \sim \varphi_R$. In \cref{fig:plane} we show the predicted relic abundance in the $m_\phi-A$ plane vs existing constraints, for the scalar mass range for which the VEV misalignment applies (we assume $T_{\rm osc} < 100$ GeV).

We summarize the scalings of the relic density and $T_{\rm osc}$ with $A$ and $m_\phi$ for the thermal and VEV misalignment regimes in the diagram below,
\begin{equation*}
\begin{gathered}
\begin{tikzpicture}[line width=1.5 pt,node distance=1 cm and 1 cm]
\filldraw[
pattern=north west lines,draw=none] (1,-0.2) rectangle (5,0.2);
\draw[thick,->] (-3.33,0) -- (9.66,0);
\node(A) at (-1.5,0.4){$T_{\rm osc} \sim m_\phi^{1/2}$};
\node (AB) at (-1, -0.4) {$\Omega_{\phi} h^2 \sim A^2 m_\phi^{-7/2}$};
\node (B) at (1,0.5) {$T_{\rm osc} \ll T_{\rm EW}$};
\node (Bo) at (1,0) {$|$};
\node (BC) at (3, -0.4) {$T_{\rm osc} \sim T_{\rm EW}$};
\node (CD) at (7.5, 0.4) {$T_{\rm osc} \sim m_\phi^{3/4}$};
\node (C) at (5, 0.5) {$T_{\rm osc} > T_{\rm EW}$};
\node (CDd) at (7, -0.4) {$\Omega_\phi h^2 \sim A^2 m_\phi^{-11/4}$};
\node (Co) at (5, 0) {$|$};
\node (q) at (9.5,0.25){~$m_\phi$};
\end{tikzpicture}
\end{gathered}
\end{equation*}
As we discuss in \cref{RHN-search} constraints from big bang nucleosynthesis (BBN) disfavor $m_\phi \lesssim 3~{\rm \mu eV}$ because this predicts \HNLs that are too light, and therefore long-lived. This constraint is relaxed somewhat if one considers larger mixing angles, as may appear for an inverse seesaw scenario. 

\subsection{Forced resonances and the electroweak phase transition \label{Forced-Res} } 
For $T_{\rm osc} \sim T_{\rm EW}$ the sudden onset of the Higgs vev can lead to forced resonances, as has been discussed in detail  in \cite{Batell:2022qvr}. When the potential minimum shifts rapidly and amplitudes are not too large, the field receives a kick which can either amplify or damp the scalar oscillations. Which of these two scenarios occurs depends both on initial conditions {\it and} on the details of the electroweak phase transition.\footnote{We thank Michael Ratz for pointing this out to us.}  

In \cite{Batell:2022qvr} the authors use a Higgs profile which has a discontinuous second derivative corresponding to a second order phase transition. In reality the electroweak ``phase transition'' is, in fact, a smooth crossover with a characteristic width of order $15~{\rm GeV}$~\cite{DOnofrio:2015gop}. We have modeled the crossover by smoothing the transition with a $\tanh(\frac{T-T_{\rm EW}}{\sigma})$ profile varying $\sigma$ from $5~{\rm GeV}$ to $45~{\rm GeV}$. We find that the details of the oscillations depend somewhat on both initial conditions and on the dynamics of the Higgs near $T_{\rm EW}$. A detailed study of forced oscillations is beyond the scope of this work, and a qualitative discussion can be found in \cite{Batell:2022qvr}. We simply note that forced oscillation phenomena may enhance, or suppress, the dark matter relic abundance for a fixed value of $A$ by a factor of roughly $\sim 5$ at most. Since $\Omega_\phi h^2 \propto A^2$ this corresponds to shifting $A$ by a factor of $2$ or so.

\subsection{Thermal mass from \HNLs \label{Thermal-HNLs} } 
When the \HNLs thermalize, they contribute to the effective potential for $\phi$ by supplying a thermal mass. Using the high temperature expansion of the fermionic thermal function $J_F$, the thermal mass is proportional to $6\mN^2(\varphi) T^2/48= g^2T^2\varphi^2/8 $. The masses of the \HNLs we consider here are sufficiently low that if a type-I seesaw mechanism is responsible for neutrino masses then \HNL thermalization takes place an order of magnitude below the electroweak scale. At these temperatures the Higgs vev has already saturated to its zero temperature value. We may check if the thermal mass from the \HNLs substantially modifies the minimum of the effective potential by checking if the following ratio\footnote{This can be seen by solving the equation $a +2b x + 4c x^3=0$ perturbatively in $b$.  One immediately finds that the relevant dimensionless ratio is $b/(a^2 c)^{1/3}$.} is $O(1)$ or larger, 
\begin{equation}
   \frac{\tfrac1{8}g^2 T^2 }{\sqrt[3]{A^2 \vEW^4\tfrac{11-3{\sf L}}{32\pi^2} g^4}} \approx  0.34 ~\frac{m_\phi T^2}{A \vEW^2} ~,
\end{equation}
for $T\lesssim  \vEW $, and where we have set ${\sf L}=0$ for simplicity and used $\varphi_0$ as an estimate for $\varphi$.  Depending on the region of parameter space we consider this may either substantially, or only slightly shift the position of the minimum. 

Notice, however, that since this all occurs at times after the electroweak phase transition, and $T_{\rm osc} \gg T_{\rm EW}$ (for thermal misalignment\footnote{For vev misalignment where $T_{\rm osc} \sim 10~{\rm GeV}$ the thermal contribution from $N$ could modify the misalingment, but we do not consider this here.}), the field is already completing many oscillations each Hubble time. The onset of the thermal population of \HNLs therefore appears adiabatic to the oscillating scalar field in the same way that the shifting minimum at higher temperatures and from the onset of the Higgs vev both appear adiabatic. The effect of the \HNLs therefore only serves to shift the vev of the scalar field, but does not substantially affect the amplitude of oscillations and therefore the dark matter relic abundance. This will lead to a slight time dependence in the mass of the \HNLs \!\!, but we do not expect this to have any sizeable cosmological consequences. Although we do not pursue this issue further here, it may be interesting to consider this effect in more detail. 


%
\subsection{Parametric resonances \label{Resonances} }
Scalar fields oscillating in a non-linear potential can produce scalar quanta. At sufficiently large occupation numbers, Bose enhancements can lead to exponential growth and this can destabilize the dynamics of the zero mode \cite{Kofman:1994rk}. For a scalar field coupled to fermions as we consider here, fermion production is also possible \cite{Greene:2000ew}. Since the dynamics here involve $|\varphi(t)- \varphi_{\rm min}| \ll \varphi_{\rm min}$ the fermion mass is approximately constant, and much heavier than the effective scalar mass at a given temperature $\mN \gg \mu_\phi$. We therefore focus on parametric resonances involving the scalar field itself. 

All oscillations occur close to the  temperature dependent minimum of the potential $\varphi_{{\rm min},T}$ such that anharmonic effects are suppressed. After the onset of oscillations we will have $\mu_\phi(T) \gg H$, and since $\varphi_{\rm min}(T) \propto T^{2/3}$ varies adiabatically with respect to the dynamics of the oscillations, it is therefore legitimate to treat $\mu_\phi$ and $H$ as constant. This is equivalent to a multi-scale separation of the field into its slow and fast modes. We will be interested in the parametric resonances of the fast modes, as a function of the wavenumber $\vb{k}$, the temperature $T$, and the effective mass $\mu_\phi(T)$. 

With this adiabatic approximation in mind we will treat the temperature $T$ as a label, such that we may consider the dynamics of $\phi_T(x,t)$. We will separate the field into its temperature dependent minimum (which does not depend on $t$ in the adiabatic approximation) and fluctuations about said minimum, $\phi_T(x,t)=\varphi_{\rm min}(T) + \delta \phi_T(x,t)$. When expanded about the minimum, the potential will have a Taylor series given by 
\begin{equation}
    V(\varphi_{\rm min} + \delta\phi_T) = V(\varphi_{\rm min})+  \frac{1}{2}\mu_\phi^2 \delta\phi_T^2 + \frac1{3!} \kappa_3 \delta\phi_T^3 +\frac1{4!} \kappa_4 \delta\phi_T^4 + ... 
\end{equation}
where $\left. \kappa_n \equiv \partial^n V / \partial \phi_T^n \right|_{\phi_T = \varphi_{\rm min}}$. The equations of motion for the fluctuating field are given by 
\begin{equation}
    \Box \delta\phi_T(x,t) + 3 H \partial_t \delta\phi_T(x,t) + V' =0 ~, 
\end{equation}
where the prime denotes differentiation with respect to $\delta \phi_T$. Let us further split the field $\delta\phi_T$ into its $\vb{k}=0$ mode, and its $\vb{k}\neq0$ modes (labeled with comoving momenta)
\begin{equation}
    \delta\phi_T(x,t) = \delta\varphi_T(t) + \sum_{k\neq0} \e^{\iu \vb{k} \cdot \vb{x} } \delta\phi_{\vb{k}}(t)~.
\end{equation} 
We will study the linearized equations of motion for $\phi_{\vb{k}}(t)$, which are given by
\begin{equation}
    \dv[2]{t} \delta\phi_{\vb k} + 3 H \dv{t} \delta\phi_{\vb k} + \bigg[\frac{\vb{k}^2}{a^2}+ \mu_\phi^2\bigg]\delta\phi_{\vb k} + \bigg[\sum_{n\geq3} \frac{\kappa_n }{(n-2)!}\qty(\delta\varphi_T )^{n-2} \bigg]\delta\phi_{\vb{k}} =0~,
\end{equation}
where $a$ is the scale factor.
This may be re-written as 
\begin{equation}
    \dv[2]{t} \delta\phi_{\vb k} + 3 H(T) \dv{t} \delta\phi_{\vb k} + \Omega^2_T(\vb{k})\delta\phi_{\vb k} + f_T(t) \delta\phi_{\vb{k}} =0~,
\end{equation}
where the definition of each variable is clear by comparison with the equation above.

Close to $T=T_{\rm osc}$ the damping due to Hubble friction occurs on time scales comparable to the oscillations. We do not, therefore, expect any periodic drive in this regime. Although it is possible that some fraction of the energy in the $k=0$ mode may leak into other modes, no exponential growth of perturbations will occur. At low temperatures $T\ll T_{\rm osc}$ we have $\mu_\phi \gg 3 H$ and Hubble friction may be treated as a small perturbation. In this limit the field will have red-shifted and its amplitude correspondingly decreased such that 
\begin{equation}
    f_T(t) \approx \kappa_3 \delta\varphi_T(t) \approx  \kappa_3 \mathcal{A}_T \cos(\mu_\phi t)~,
\end{equation}
where $\kappa_3 = \tfrac23 \mu_\phi^2/\varphi_{\rm min}(T)$ for $L_{\rm osc} \sim 1.5$. We have also introduced the temperature dependent amplitude 
\begin{equation}
    \mathcal{A}_T = 0.145 \times \left(\frac{\mu_\phi(T_{\rm osc})}{\mu_\phi(T)}\right)^{1/2} \left(\frac{g_*(T)}{g_*(T_{\rm osc})}\right)^{1/2} \varphi_{\rm min}(T_{\rm osc})  \qty(\frac{T}{T_{\rm osc}})^{3/2} ~. 
\end{equation}
In this limit we can map onto the Mathieu equation \cite{NIST:DLMF}, 
\begin{equation}
    \ddot{y} + \nu^2 \big( 1+ h \cos\omega t \big)      y= 0~ . 
\end{equation}
The mapping between our problem and the parameters of the Mathieu equation is given by $\nu^2=\mu_\phi^2 + (k/a)^2$ and $h = \kappa_3 \, {\cal A}_T / \nu^2$.

In the limit of $h\rightarrow0$ the condition for a parametric resonance may be written as \cite{Landau:1991aaa}
\begin{equation}
    \label{par-res-cond}
    |2\nu-\omega | < \tfrac12|\nu h|~ \implies~ 2\sqrt{\mu_\phi^2+\frac{\mathbf{k}^2}{a^2}}  -\mu_\phi<\frac{ \mu_\phi^2 \mathcal{A}_T}{3 | \varphi_{\rm min}(T)|\sqrt{\mu_\phi^2+\frac{\mathbf{k}^2}{a^2}} }~,
\end{equation}
which may be re-expressed as 
\begin{equation}
 \abs{\frac{\mathcal{A}_T}{3\varphi_{\rm min}(T)}} > \left(2 \sqrt{\frac{\mathbf{k}^2}{a^2 \, \mu_\phi^2}+1} -1 \right) \sqrt{1+\frac{\mathbf{k}^2}{a^2 \, \mu_\phi^2}} \gtrsim 1 ~.
\end{equation}
Therefore, provided $\mathcal{A}_T< 3\varphi_{\rm min}(T)$ no parametric resonance occurs at this order in the linearized equations of motion; this inequality is always satisfied in practice. We therefore find that parametric resonances are unimportant for the dynamics of the ultralight dark matter.

Higher order terms from the potential can induce parametric resonances via higher harmonics of the field. For example the quartic term, proportional to $\cos^2(\mu_\phi t)= \tfrac12(1+ \cos2 \mu_\phi t)$, both provides a parametric oscillation {\it and} a detuning of the natural oscillator's frequency. This contribution is further suppressed by $\mathcal{A}_T/\varphi_{\rm min}(T)$ (in addition to a typically small logarithm) and is negligible relative to the term proportional to $\cos(\mu_\phi t)$. We therefore conclude that parametric resonances can be neglected when estimating the dark matter relic abundance.

\subsection{Initial conditions after inflation \label{Preheating} }
Thermal misalignment erases initial conditions much smaller than $\varphi_{\rm pre}(T_{\rm osc}) \sim \varphi_0$, which for the parameter space we consider is typically of order $\sim 10^{13}~{\rm GeV}$. It remains possible that certain inflationary scenarios may result in a much larger initial misalignment, in which case the dark matter relic abundance cannot be predicted given only $A$ and $m_\phi$. Although we remain agnostic about the details of inflation, it is interesting to speculate on what kinds of initial conditions are generic in the model we consider here. In contrast to the quadratic potential considered in \cite{Batell:2022qvr},  the nonlinearities in the CW potential, and the dependence of $\mN$ on $\varphi(t)$, mean that large field velocities can result in non-perturbative particle production \cite{Kofman:1997yn,Felder:1998vq,Greene:1998nh,Greene:2000ew}. This offers an efficient damping mechanism, sometimes referred to as instant preheating \cite{Felder:1998vq}.

If we consider inflationary scenarios with a low Hubble scale, $H_I \ll T_{\rm EW}$, then the analysis sketched in \cite{Batell:2022qvr} applies. Electroweak symmetry is broken during inflation, and field values diffuse towards $\varphi_0$ \cite{Starobinsky:1994bd,Graham:2018jyp}. They remain ``pinned'' there by Hubble friction, and large values of $A$ are required to realize a phenomenologically viable dark matter relic abundance. 

It is arguably more natural to consider the opposite limit where $H_I \gg T_{\rm EW}$. In this limit one expects larger field values of order $\varphi_* \sim H_I/g$ \cite{Starobinsky:1994bd,Graham:2018jyp}. This will lead to a large effective mass $m_{\rm eff}^2 \sim g^2 H_I^2$.   After the end of inflation, assuming radiation domination, Hubble will drop like $H\sim T^2$, and the field will begin oscillating at  temperature $T^*_{\rm osc} \sim \sqrt{g M_{\rm Pl} H_I}$. This is much larger than the $T_{\rm osc}$ relevant for thermal misalignment for $H_I \gg T_{\rm EW}$. The field will loose energy via particle production, and its amplitude will decrease. We may characterize its motion by the field value at each successive turning $\varphi_{\rm turn}^{(i)}$ point where $\dot{\varphi}=0$; energy loss implies $\varphi_{\rm turn}^{(i)} < \varphi_{\rm turn}^{(i-1)}$. Since $m_{\rm eff}^2$ is field-dependent, and smallest near $\varphi=0$, it is generic that the field will get stuck by Hubble friction close to the origin in field space. This then suggests $\varphi_I \approx 0$ as a ``natural'' initial condition for inflationary scenarios satisfying $H_I \gg T_{\rm EW}$. In practice an equivalent condition is $\varphi_I \ll 10^{13}~{\rm GeV}$ due to the erasure of initial conditions by thermal misalignment. 

It is interesting to note that the ability to accommodate large field values during inflation naturally suppresses isocurvature fluctuations \cite{Piazza:2010ye}. The relaxation mechanism sketched above may therefore allow the model we consider to both evade isocurvature constraints while simultaneously having small enough initial conditions to allow for erasure via thermal misalignment.

\subsection{Connections to leptogenesis \label{Leptogenesis}}
Before moving on to phenomenological signatures, let us comment on the possibility of explaining the baryon asymmetry within the model. We have so far remained entirely agnostic as to the Yukawa couplings with charged leptons that generate neutrino masses. Scalar masses above $30~{\rm \mu eV}$ are both consistent with constraints from BBN, and predict \HNLs in the mass range of a few GeV. It is well known that leptogenesis by oscillations (or ARS leptogenesis for Akhmedov, Rubakov, and Smirnov~\cite{Akhmedov:1998qx}) is operational in this mass range. It is therefore tempting to ask if the mechanism may operate within the model at hand, thereby supplying an explanation of dark matter, neutrino masses, and the observed baryon asymmetry. 

Indeed we expect, given the extremely small Yukawa coupling between $N$ and $\phi$, that the lifetime of the \HNLs will be largely unaffected by the scalar field. We note, however, that close to $T_{\rm osc}$ the \HNL mass will fluctuate an $O(1)$ amount, albeit with a very slow period of $\mu(T_{\rm osc})$. It may be possible for this to lead to the production of \HNLs modifying the non-equilibrium number densities of $N$ with respect to a vanilla ARS leptogenesis mechanism. This may allow for efficient leptogenesis at smaller mixing angles. A detailed investigation of leptogenesis within this model lies beyond the scope of this paper. We note that there is a large volume of parameter space where the mechanism is viable \cite{Klaric:2021cpi}. 

\section{Experimental signatures and constraints \label{Pheno}}
One interesting consequence of the radiatively generated parameter space we consider here is the correlation between \HNL phenomenology and direct searches for a light scalar. At higher masses, where thermal misalignment is operational, inverse square law tests may offer the most competitive search channel with which to probe the parameter space that predicts the correct relic abundance. At lower masses, where $A$ may be very small and still produce acceptable misalignment to supply the correct dark matter abundance, direct searches for the Higgs portal coupling may be unrealistic. In this limit searches for the \HNLs which radiatively generate the light scalar's mass may offer better experimental prospects. 

\subsection{Probes of a light scalar \label{Yukawa-search} } 
Direct searches for the light mediator $\phi$ have been discussed previously in the literature, and we refer the interested reader to Refs.\ \cite{Piazza:2010ye,Batell:2022qvr} for a more detailed discussion. For masses between $3~{\rm \mu eV}$ and $15~{\rm meV}$ the strongest constraints come from short-distance tests of the inverse square law (on the scale of millimeters). The scalar discussed here will result in a Yukawa potential with a range of $\lambda= 1/m_\phi = 1~{\rm mm} \times \qty(\frac{0.197~{\rm meV}}{m_\phi})$. The strength of the coupling is set by the nucleon-scalar coupling discussed in \cref{nucleon-scalar}.

Inverse square law tests are often quoted in terms of $\alpha^2= \alpha_1 \alpha_2$ where a potential of the form $V=-Gm_{\rm nuc}^2/r \times (1+ \alpha^2 \e^{-m_\phi r})$ is assumed, being $m_{\rm nuc}$ the nucleon mass. The Yukawa potential's strength,  $\alpha^2$, is related to the Higgs portal coupling, $A$, via 
\begin{equation}
    \alpha = \frac{\Lambda_{\rm had}}{m_h ^2} \frac{\sqrt2 ~M_{\rm Pl}}{m_{\rm nuc}} \times A ~,
\end{equation}
where $\Lambda_{\rm had}$ is a hadronic scale of order $ \sim 200~{\rm MeV}-600~{\rm MeV}$ \cite{Shifman:1978zn,Cheng:1988cz,Barbieri:1988ct} and $m_h=125~{\rm GeV}$ is the mass of the Higgs; for numerical estimates we take $\Lambda_{\rm had} =530~{\rm MeV}$ following \cite{Cheng:1988cz,Barbieri:1988ct}.

A number of groups have obtained limits on a Yukawa potential at millimeter scales. Relevant experimental tests include the Irvine \cite{PhysRevD.32.3084}, E\"ot-Wash \cite{PhysRevLett.98.021101,PhysRevLett.124.101101}, and HUST experiments \cite{PhysRevLett.124.051301}. Projections from the HUST group predict improved sensitivity in the vicinity of $m_\phi \sim {\rm meV}$, and will test part of the relic-abundance parameter space for $m_\phi \sim 1~{\rm meV}$ \cite{Du:2022veu}.

\subsection{Right-handed neutrino searches \label{RHN-search} } 
Fixing the relic density of $\phi$ to be all the dark matter predicts the relationship between \HNL mass and $m_\phi$. For $m_\phi \gtrsim 0.1~{\rm meV}$, where $T\gtrsim T_{\rm EW}$ we find 
\begin{equation}
    \mN = 6.1~{\rm GeV} ~\qty(\frac{m_\phi}{3~\rm meV})^{3/16} ~,
\end{equation}
which is relatively insensitive to the dark matter mass $m_\phi$. More generally $\mN\sim (A/m_\phi)^{1/2}$ as given by \cref{RHN-mass}. If $T \ll T_{\rm EW}$ instead, or in terms of the scalar mass, $m_\phi \ll 0.1~{\rm meV}$, the mass of the heaviest neutrino is instead given by 
\begin{equation}
    \mN = 1.25 ~{\rm GeV} ~\qty(\frac{m_\phi}{90~\rm \mu eV})^{3/8} ~.
\end{equation}
Constraints from BBN suggest a minimal mass for \HNLs \!\!. This constraint is set by the lifetime of the \HNL \!\!, and so depends also on the Yukawa couplings. If we fix $Y\sim \sqrt{m_\nu \mN}/\vEW$ to the seesaw line and demand $\tau_N \lesssim 0.1~{\rm s}$ \cite{Ruchayskiy:2012si} then one has $\mN \gtrsim 1~{\rm GeV}$. This then demands that $m_\phi \gtrsim 30 ~{\mu \rm eV}$ to avoid standard constraints on \HNLs from BBN (see \cref{fig:plane}).

The radiative scenario presented above can be studied either by direct probes of a light scalar (e.g.\ fifth force searches)  or by searching for the \HNLs that are predicted in the spectrum. As is well appreciated in the literature surrounding \HNLs, mixing angles can be much larger than a naive type-I seesaw estimate would suggest (e.g.\ if the inverse seesaw mechanism is operational \cite{Mohapatra:1986aw}). Therefore, one may view searches for \HNLs below a few GeV as generic {\it discovery} opportunities for the dynamics discussed herein. 

Searches for \HNLs can also exclude certain regions of viable dark matter parameter space if the seesaw line can be probed. No near-term experiment projects sensitivity down to the seesaw line for masses that are not excluded from BBN, however SHiP projects sensitivity to $\theta_\mu^2$ and $\theta_e^2$ that is within an order of magnitude of the seesaw line \cite{SHiP:2018xqw}.

If one is willing to entertain non-standard cosmologies, or alternative decay paths for the \HNLs, then the bound on the scalar mass discussed above is relaxed. A few near-term experiments then offer promising  detection prospects. For $140~{\rm MeV}\lesssim \mN \lesssim460~{\rm MeV}$ NA62 has probed $\theta_\mu^2$ to within an order of magnitude of the seesaw line \cite{NA62:2020mcv}, and may reach it with further data. For $60~{\rm MeV} \lesssim \mN \lesssim 130~{\rm MeV}$ the upcoming PIONEER experiment projects to probe the seesaw line for $\theta_{e}^2$ \cite{PIONEER:2022yag}. For $\mN \lesssim 1~{\rm MeV}$  the \HNLs can be searched for in nuclear beta decays.  BeEST \cite{Leach:2021bvh}, HUNTER \cite{Martoff:2021vxp}, and KATRIN \cite{KATRIN:2022spi} project sensitivity below the seesaw line \cite{Acero:2022wqg}, and may offer a complimentary discovery avenue.

\section{Conclusion and Outlook \label{Conclusions} }
We have considered ultralight Higgs-portal dark matter in the presence of a neutrino mass mechanism. Since the scalar must be a gauge singlet in order to couple via the super-renormalizable Higgs portal, it is generic to consider interactions between scalar dark matter and singlet fields responsible for neutrino mass generation. In our case, we focus on \HNLs and a type-I seesaw mechanism. Generically, \HNLs much heavier than the scalar dark matter, $\mN \gg m_\phi$, induce a large radiative mass for $\phi$.

Motivated by this observation, we have focused on a region of parameter space in which the scalar's mass is {\it entirely} generated by radiative effects. In this CW-dominated regime the model has correlated and testable predictions, with phenomenology controlled by two parameters (plus Yukawa couplings to fix neutrino masses). From a microphysical perspective, the two input parameters are the \HNL scalar Yukawa coupling $g$, and the soft Higgs portal coupling $A$. We choose to trade $g$ for the physical scalar mass at zero temperature $m_\phi$. The mass scale of \HNLs is predicted by $A$ and $m_\phi$, and there exists a preferred range for dark matter between a few ${\mu\rm eV}$ and a few ${\rm meV}$. \HNL searches offer a complimentary discovery avenue in addition to direct probes of an ultralight scalar via inverse square law tests. 

While the CW-dominated parameter space we consider here is highly predictive, and represents some $O(1)$ fraction of the available parameter space,\footnote{Quantifying volumes in parameter space with a log-measure.} it is by no means generic. The parameter space of an ultralight singlet scalar dark matter in the presence of a neutrino mass mechanism is worth exploring more broadly. The scalar will generically couple to whatever UV degrees of freedom generate neutrino masses and so the dynamics of dark matter will be tied to the physics of neutrino masses in a manner similar to what is presented above. We plan to pursue this connection in future work.

\section*{Acknowledgments}
We thank  Akshay Ghalsasi and especially Brian Batell for useful discussions about thermal misalignment in the context of the Higgs portal. Mark Wise supplied helpful comments on the possibility of parametric resonances. We acknowledge useful discussions with Junwu Huang, Maxim Pospelov, Marilena Loverde, Michael Ratz, and Yohei Ema about initial conditions after inflation. We thank Brian Batell, Akshay Ghalsasi, Peizhi Du, Michael Ratz, and Mark Wise for feedback on the manuscript.  We acknowledge the hospitality of the Simons Center for Physics and Geometry and the Aspen Center for Physics (which is supported by National Science Foundation grant PHY-2210452) while part of this work was being completed.

\paragraph{Funding information.}
 This work is supported by the U.S. Department of Energy, Office of Science, Office of High Energy Physics, under Award Number DE-SC0011632 and by the Walter Burke Institute for Theoretical Physics. RP is supported by the Neutrino Theory Network under Award Number DEAC02-07CHI11359, the U.S. Department of Energy, Office of Science, Office of High Energy Physics, under Award Number DE-SC0011632, and by the Walter Burke Institute for Theoretical Physics.

\appendix

\section{Estimate of radiatively generated terms \label{app:rad_corr_est}}

Here we present estimates of the radiatively generated coefficients that are induced in the scalar potential (see Eq.~\eqref{model-def}). These are dominated by the $g \phi N^c N^c$ contribution. 
\begin{align}
    c_4    &\sim  \frac{g^4}{16\pi^2}     ~,\\ 
    B      &\sim  \frac{Y^2 g^2}{16\pi^2} ~.
\end{align}
The cubic term $c_3\phi^3$, which does not respect the $\mathbb{Z}_4$ symmetry, is induced by the soft-breaking term $A \phi |H|^2$ through the mass of the RHN,
\begin{align}
      c_3 &\sim \frac{g^3 m_N }{16\pi^2} \sim \frac{g^4 \varphi_0}{16\pi^2} \sim \frac{g^{8/3} (A \vEW^2)^{1/3}}{16\pi^2}~.
\end{align}
Although the soft-breaking term contributes directly to all the above coefficients, $c_4 \sim A^4/(16\pi^2 m_h^4)$, $B \sim A^2 / (16 \pi^2 m_h^2)$ and $c_3 \sim A^3/(16\pi^2 m_h^2)$, these are subdominant for the parameter space consistent with the observed relic density. We do not discuss $c_2$ or $M_N$ since these are explicitly discussed in the main text. 

\section{Fine-tuning}
In the main text we have emphasized that the model under consideration does not suffer from a fine-tuning problem when considered as an effective theory valid below the electroweak scale. Since the notion of fine-tuning can mean different things to different audiences, we clarify our precise meaning in this appendix. 

To illustrate the problem we have in mind, let us consider $B\sim O(1)$ in Eq.~\eqref{model-def}. In this case, a loop involving the Standard Model Higgs contributes a mass to the scalar that scales as $c_2 \sim B m_h^2$. A renormalization condition can be chosen such that $c_2 \ll m_h$ at some scale $\mu\sim m_\phi$. However, at different subtraction points, the strong renormalization group flow of the parameter $c_2$ will guarantee that $c_2 \sim m_h$ at other renormalization scales. This  complicates the construction of a perturbation theory, and raises questions about whether or not the parameter space considered is stable against radiative corrections without severe fine-tuning. This notion of fine-tuning is what motivates us to consider incredibly small values of $B \ll m_\phi^2/m_h^2$. A similar problem emerges from loops of \HNLs. Nevertheless, since most of the Lagrangian parameters are taken to be {\it generated} by radiative corrections, they will be radiatively stable i.e., ``of their natural size''. 

We contrast the above notion of fine-tuning, which is defined at the level of an effective theory applicable at and below the weak scale, with the notion of ``UV naturalness''. This latter notion would estimate the Higgs loop mentioned above as providing a contribution of order $c_2 \sim B \Lambda_{\rm UV}^2$, where $\Lambda_{\rm UV}$ is a ``high energy scale'' which is often taken to be the Planck mass. With this notion of naturalness, the Higgs mass has a naturalness problem. Note, however, that it does not suffer from fine-tuned radiative corrections within the low-energy effective theory \cite{Wells:2021zdp}. 

The notion of UV naturalness is essentially a problem of wedding the Standard Model with generic UV completions. It is of course interesting to ask if demanding that such UV completions are themselves natural in the sense of renormalization group flows can supply interesting lampposts for future experimental discovery. As is well appreciated, the solution to the Higgs ``naturalness problem'' is unsolved, and we do not know what the correct description of nature is above the electroweak scale. We do not claim to have solved this problem, but only to have studied a model that is (as emphasized in the main text) {\it i)} radiatively stable below the electroweak scale, {\it ii)} capable of explaining the dark matter relic abundance and neutrino masses, and {\it iii)} where these explanations can be linked via the model's radiative (i.e., Coleman-Weinberg) structure. 
%

\bibliographystyle{jhep}
\bibliography{bib_plurgui.bib}

\end{document}